\documentclass[aps,prb,twocolumn,reprint,floatfix,superscriptaddress]{revtex4-2}

\usepackage[english]{babel}
\usepackage[T1]{fontenc}
\usepackage[utf8]{inputenc}
\usepackage{graphicx, amsmath, amssymb, dsfont, physics, xcolor, booktabs}
\graphicspath{{figures/}}

\definecolor{myblue}{rgb}{0.2,0.2,0.8}
\definecolor{myred}{rgb}{1,0.,0.3}

\usepackage[colorlinks=true,citecolor=myblue,linkcolor=myred]{hyperref}

\newcommand{\ueV}{\;\mu\text{eV}}

\begin{document}

\title{Long-range spin transport in asymmetric quadruple quantum dots configurations}

\author{David Fernández-Fernández}
\email{david.fernandez@csic.es}
\affiliation{Instituto de Ciencia de Materiales de Madrid ICMM-CSIC, 28049 Madrid, Spain.}
\author{Johannes C.~Bayer}
\affiliation{Institut für Festkörperphysik, Leibniz Universität Hannover, D-30167 Hanover, Germany}
\affiliation{Physikalisch-Technische Bundesanstalt, D-38116 Braunschweig, Germany}
\author{Rolf J.~Haug}
\affiliation{Institut für Festkörperphysik, Leibniz Universität Hannover, D-30167 Hanover, Germany}
\author{Gloria Platero}
\affiliation{Instituto de Ciencia de Materiales de Madrid ICMM-CSIC, 28049 Madrid, Spain.}
\date{\today}

\begin{abstract}
	We theoretically investigate long-range coherent charge transport in linear quadruple quantum dot (QQD) arrays under reduced symmetry configurations.
	Employing a master equation approach, we identify precise resonant conditions that enable minimal occupation of intermediate dots, thereby facilitating long-range transfer between distant sites.
	Our results highlight the critical role of parameter asymmetry and coherent tunneling mechanisms in achieving efficient quantum state transfer.
\end{abstract}

\maketitle

\section{Introduction}
Quantum dots (QDs) constitute a highly controllable platform for exploring fundamental quantum phenomena, coherent charge transport, and potential applications in quantum information processing and quantum computing architectures \cite{Taylor2005, Burkard2023}.
Among various configurations, linear quantum dot arrays have emerged as especially promising, owing to their ability to support long-range coherent charge and spin transport mediated by virtual occupation of intermediate dots, a process known as co-tunneling \cite{DeFranceschi2001,Shinkai2009,Weymann2011,Braakman2013}.

Substantial insight into coherent transport mechanisms has been obtained from studies on double quantum dot (DQD) \cite{Wiel2002,Jouravlev2006,Busl2010a,Wang2016,Dani2022,FernandezFernandez2023} and triple quantum dot (TQD) systems \cite{Saraga2003,Gaudreau2006,Michaelis2006,Schroeer2007,Rogge2008,Rogge2009,Granger2010,Villavicencio2011,Gaudreau2011,Amaha2012,Amaha2013,Kotzian2016,Wang2016a,PicoCortes2019,Hendrickx2020}.
In DQDs, spin blockade phenomena, arising from Pauli exclusion, have been extensively studied and exploited for spin qubit readout and coherent manipulation \cite{Horstig2024}.
TQD systems have revealed more intricate phenomena, including bipolar spin blockade \cite{Busl2013}, coherent superpositions avoiding intermediate dots \cite{Sanchez2014}, and long-range spin and charge transfer mediated by superexchange interactions \cite{Sanchez2014a}, where the transfer occurs between not directly coupled QDs.
These seminal investigations established the pivotal role of coherent tunneling and spin-dependent interactions in quantum dot arrays.

Extending these concepts to quadruple quantum dot (QQD) systems \cite{Haug1992,Thalineau2012,Delbecq2014,Fujita2017,Bayer2017,Bayer2019,Ban2019,Kandel2019,Mills2019} introduces additional complexity and richness due to the increased number of degrees of freedom and an expanded parameter space.
As an example, Kondo effect \cite{Liu2010} and Nagaoka's ferromagnetism can be observed in QQDs \cite{Stecher2010}.
QQDs have been also proposed as quantum simulators of the Wigner molecule \cite{Creffield2002}, rendering this system particularly interesting for quantum simulation and quantum information applications.
Nevertheless, achieving coherent long-range transfer in QQDs it is expected to require highly symmetric configurations, which are difficult to obtain in experiments.
This difficulty stems from the demanding tuning of requirements for multiple parameters, including tunneling amplitudes, inter-dot Coulomb interactions, and gate-defined energy levels.
Currently, assisted machine learning techniques are proposed to simplify the tuning of these parameters \cite{Kalantre2019}.
However, the existence of long-range transport in reduced symmetric configurations remains an open question.

\begin{figure}[t!]
	\centering
	\includegraphics[width=\linewidth]{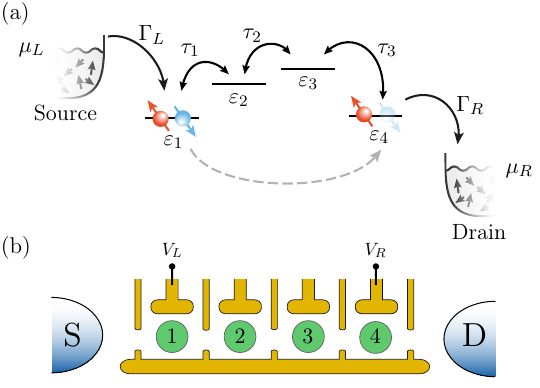}
	\caption{(a) Linear quadruple quantum dot array, coupled to leads with rates $\Gamma_{L,R}$, defined by chemical potentials $\mu_{L,R}$.
	Electrons can flow from the source (left lead) to the drain (right lead) through the quantum dot array.
	Interdot tunneling rates are denoted by $\tau_i$.
	Each dot is characterized by its on-site energy $\varepsilon_i$.
	When states $(2, 0, 0, 1)$ and $(1, 0, 0, 2)$ are degenerate, and the middle dots are out of resonance, long-range transport occurs between the first and last dots, denoted with a gray dashed line.
	(b) Schematic representation of gates defining the QQD, together with the source and drain leads.
	The plunger states $V_L$ and $V_R$ are applied to the first and fourth dots, respectively.}
	\label{fig:schematic}
\end{figure}

In this work, we present a theoretical study of charge transport in a linear QQD array coupled to electronic reservoirs where the symmetry requirements are relaxed.
Then, we focus on coherent long-range charge transfer under explicitly non-symmetric configurations.
Using a master equation formalism in conjunction with an effective co-tunneling description, we systematically explore the parameter regimes that enable long-range transport between distant quantum dots.
We identify distinct resonant conditions characterized by suppressed occupation of the intermediate sites.

\section{Model}
We consider a minimal model comprising Anderson impurities linearly coupled to two fermionic leads, see Fig.~\ref{fig:schematic}~(a), described by the total Hamiltonian $\hat{H} = \hat{H}_0+\hat{H}_{\text{leads}}+\hat{H}_{\text{coup}}$.
The first term describes the quantum dot array populated by electrons, with a single orbital per site.
This term is modeled by a Fermi-Hubbard Hamiltonian, given by
\begin{align}
	\begin{split}
		\hat{H}_0 = & \sum_{i=1}^N\varepsilon_i \hat{n}_i + \sum_{i=1}^N U_{i}\hat{n}_{i\uparrow}\hat{n}_{i\downarrow} +\sum_{i\neq j}^N \frac{V_{i, j}}{2}\hat{n}_{i}\hat{n}_{j} \\
		            & +\sum_{i=1}^{N-1}\sum_{\sigma=\left\{\uparrow, \downarrow\right\}} \tau_{i} \left(\hat{c}_{i\sigma}^\dagger \hat{c}_{(i+1)\sigma} + \text{H.c.} \right),
	\end{split}
	\label{eq:hamiltonian_0}
\end{align}
where $\varepsilon_i$ denotes the on-site energy of the $i$-th dot, $U_i$ is the on-site Coulomb repulsion, $V_{i,j}$ represents the inter-dot Coulomb repulsion, and $\tau_{i}$ is the hopping amplitude between adjacent sites $i$ and $i+1$.
Note that no long-range hopping is considered in this model.
In this work, we aim to study electron spin qubits on GaAs planar quantum dots, where the spin-orbit interaction is negligible.
However, the main results of this work can be extended to other materials with negligible spin-orbit interaction.
For simplification, we assume that the inter-dot Coulomb repulsion depends only on the distance between the dots, i.e., $V_k \equiv V_{i,i+k}$.
The on-site energies are determined by the applied voltage bias.
Due to capacitive cross-talk, these energies are not independent, and are expressed as a linear combination of the voltages applied to the two outermost gates, see Fig.~\ref{fig:schematic}~(b), as
\begin{equation}
	\varepsilon_i = \varepsilon_{i0} + \alpha_{iL}V_L + \alpha_{iR}V_R,
	\label{eq:lever_arms}
\end{equation}
where $\varepsilon_{i0}$ is a static energy offset, $\alpha_{iL(R)}$ are the lever arms associated with the left (right) gate, and $V_{L(R)}$ are the corresponding voltage biases.
We set the energy offsets of the first and last dots to zero, i.e., $\varepsilon_{10}=\varepsilon_{N0}=0$.
For concreteness, we adopt the lever arm values reported in \cite{Bayer2019}: $\alpha_{iL} = (-85, -36, -12, -7)\;\mu \text{eV}/\text{mV}$ and $\alpha_{iR}=(-7, -12, -32, -81)\;\mu \text{eV}/\text{mV}$.

The leads are modeled by the Hamiltonian
\begin{equation}
	\hat{H}_{\text{leads}} = \sum_{l=L,R}\sum_k\sum_\sigma\varepsilon_{lk}\hat{d}_{lk\sigma}^\dagger \hat{d}_{lk\sigma},
\end{equation}
where $\hat{d}_{lk\sigma}^\dagger$ ($\hat{d}_{lk\sigma}$) creates (annihilates) an electron in lead $l$ with momentum $k$, spin $\sigma$, and energy $\varepsilon_{lk}$.

Finally, the coupling between the leads and the quantum dots is described by
\begin{equation}
	\hat{H}_{\text{coup}} = \sum_{l,k,i,\sigma}\left(\gamma_{l, i}\hat{d}_{lk\sigma}^\dagger \hat{c}_{i\sigma} + \text{H.c.} \right),
\end{equation}
where $\gamma_{l,i}$ quantifies the coupling strength between lead $l$ and dot $i$.
We assume a linear array of $N$ dots, with the first and last dots coupled to the left and right leads, respectively.
Thus, the nonzero couplings are $\gamma_{L,1}=\gamma_L$ and $\gamma_{R,N}=\gamma_R$, while $\gamma_{L,i}=\gamma_{R,i}=0$ for all other sites.

The charge current is obtained by solving the steady-state of the Lindblad master equation for the reduced density matrix of the dots \cite{Busl2010}:
\begin{equation}
	\dot{\hat{\rho}} = -i\left[\hat{H}_0, \hat{\rho}\right]+\mathcal{L}\hat{\rho} = 0.
\end{equation}
The Liouvillian superoperator $\mathcal{L}$ accounts for the coupling to the leads, as well as spin relaxation and dephasing processes.
A detailed derivation of the Lindblad master equation, along with the general parameters used throughout this work, is provided in Appendix~\ref{app:master_eq}.

Our primary objective is to analyze long-range transport across a quadruple quantum dot array, with electric current flowing from the left to right.
To this end, we examine the charge current as a function of the gate voltages applied to the extremes of the quantum dot array.
Additional details on the current operator can be found in Appendix~\ref{app:current_op}.

We now introduce the notation for the quantum states considered.
Focusing on the charge sector, basis states are labeled by the site occupations as $\ket{N_1N_2N_3N_4}$.
The corresponding energy are denoted $E_{N_1N_2N_3N_4}$.
Since no magnetic field is applied to the system, these energies are independent of spin.
However, due to the presence of Coulomb interaction, the eigenenergies lift the degeneracy between states with different spin configurations.

Throughout this manuscript, we consider configurations with a total electron number ranging from $0$ to $4$.
Simulations involving larger occupation numbers (from $5$ to $8$ electrons) yield qualitatively similar results and are not shown here.

In this manuscript, we focus specifically on long-range transitions between the first and last dots in the QQD array.
To facilitate this, we define a dark state $\ket{DS}$ as an eigenstate of the system Hamiltonian that exhibits zero population on the two central dots.
It is important to note that this definition of a dark state originates within the quantum state transfer framework \cite{Greentree2004,Ban2018,FernandezFernandez2024} and differs from its usage in the context of charge transport, where a dark state is typically understood as one that does not contribute to current flow \cite{Sanchez2014a,FernandezFernandez2023}.

\section{One particle results}
\subsection{Closed system}
We begin by analyzing the dynamics of a single electron in an isolated quadruple quantum dot array, i.e., with $\Gamma_{L,R}=0$.
In this regime, the spin degree of freedom can be neglected, and the Hamiltonian simplifies to
\begin{equation}
	\hat{H}_0 = \sum_{i=1}^N\varepsilon_i \hat{n}_i + \sum_{i=1}^{N-1} \tau_{i} \left(\hat{c}_{i}^\dagger \hat{c}_{(i+1)} + \text{H.c.} \right).
\end{equation}

Exact diagonalization reveals that, in contrast to the triple quantum dot case, a dark state is not present in the quadruple quantum dot array, even when all dots are in resonance, $\varepsilon_i = \varepsilon$.
That is, there exists no eigenstate of the Hamiltonian exhibiting vanishing occupation of the central dots.

Nonetheless, a high-order co-tunneling process can emerge when the central dots are strongly detuned from the outer ones, enabling direct tunneling between the first and last dots.
The occupation of the intermediate dots can be suppressed by tuning their on-site energies $\varepsilon_i$.
In the limit $\abs{\varepsilon_2},\abs{\varepsilon_3}\gg \abs{\varepsilon_1}, \abs{\varepsilon_4}, \tau_i$, a third-order Schrieffer-Wolff transformation yields an effective Hamiltonian involving only the outermost dots:
\begin{equation}
	\hat{H}_{\text{eff}} = \tilde{\varepsilon}_1\hat{n}_1 + \tilde{\varepsilon}_4\hat{n}_4 + \tau_{\text{eff}}\left(\hat{c}_1^\dagger\hat{c}_4 + \text{H.c.}\right),
	\label{eq:effective_model}
\end{equation}
where the renormalized on-site energies are $\tilde{\varepsilon}_1=\varepsilon_1+\tau_1^2/(\varepsilon_1-\varepsilon_2)$ and $\tilde{\varepsilon}_4=\varepsilon_4+\tau_3^2/(\varepsilon_4-\varepsilon_3)$, and the effective hopping amplitude is
\begin{equation}
    \tau_{\text{eff}}=\frac{\tau_1\tau_2\tau_3}{2}\left[\frac{1}{(\varepsilon_1 - \varepsilon_2)(\varepsilon_1-\varepsilon_3)}+\frac{1}{(\varepsilon_4 - \varepsilon_2)(\varepsilon_4-\varepsilon_3)}\right].
\end{equation}

\begin{figure}[t!]
	\centering
	\includegraphics[width=\linewidth]{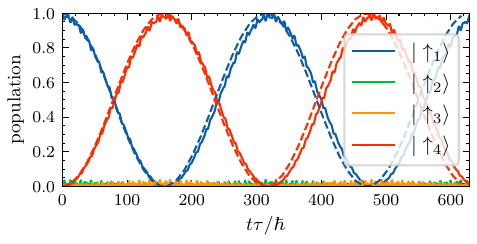}
	\caption{Dynamics (solid lines) for a single particle in a quadruple quantum dot.
		The effective model, given by Eq.~(\ref{eq:effective_model}), is shown as dashed lines.
		The system parameters are $\tau_i = \tau$, $\varepsilon_1 = \varepsilon_4=0$, and $\varepsilon_2 = \varepsilon_3 = 10\tau$.}
	\label{fig:effective_model_one_particle}
\end{figure}

When the condition $\tilde{\varepsilon}_1=\tilde{\varepsilon}_4$ is met, full charge transfer between the outermost dots becomes possible with negligible population of the central sites.
This transfer is mediated by virtual transitions to the central dots, which are energetically forbidden.
Fig.~\ref{fig:effective_model_one_particle} illustrates how the dynamics governed by the effective model closely replicate those of the full Hamiltonian, with Rabi oscillations between the first and last dots (blue and red lines), and minimal occupation of the intermediate dots (green and orange lines).

\subsection{Open system}
When the quadruple quantum dot is coupled to the electronic reservoirs, a current can flow through the system.
This occurs only when a resonant transition within the QQD is energetically allowed, and it is inside the transport window.
For a single particle, the region of positive charge current is defined by the conditions $\mu_L > \varepsilon_1$ and $\mu_R < \varepsilon_4$.
Outside this region the system resides in a Coulomb blockade regime, and the current vanishes.

Using the effective model from Eq.~(\ref{eq:effective_model}), the resonance condition $\tilde{\varepsilon}_1=\tilde{\varepsilon}_4$ results in the relation.
\begin{equation}
	\varepsilon_3 = \varepsilon_2 + \frac{\tau_3^2(\varepsilon_2-\varepsilon_1)}{\tau_1^2+(\varepsilon_1-\varepsilon_2)(\varepsilon_1-\varepsilon_4)}.
	\label{eq:resonance_condition_one_particle_detuning}
\end{equation}

It is important to remark that when the condition of large detuning between the outer and the inner dots is not fulfilled, the effective model in no longer valid.
In this case, even with $\tilde{\varepsilon}_1=\tilde{\varepsilon}_4$, the population of the middle dots remains finite.
To quantify the average charge occupation of each dot, we define
\begin{equation}
	\langle Q_i \rangle \equiv \tr(\hat{n}_i \hat{\rho}_\mathrm{ss}),
\end{equation}
where $\hat{n}_i$ is the charge occupation operator of the $i$-th dot and $\hat{\rho}_\mathrm{ss}$ is the steady-state density matrix.

\begin{figure}[t!]
	\centering
	\includegraphics[width=\linewidth]{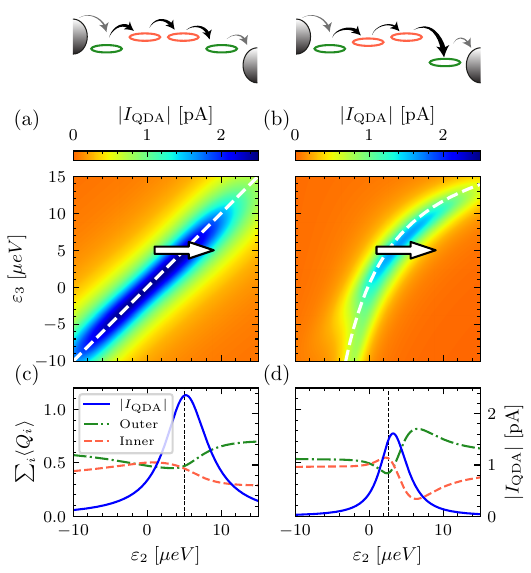}
	\caption{Comparison of current for a highly symmetric configuration (a) and an asymmetric configuration (b), as shown on the top of each panel.
	White dashed lines denote the analytical resonance condition from Eq.~(\ref{eq:resonance_condition_one_particle_detuning}).
	Panels (c, d) show the average charge (left axis) and current (right axis) along the white horizontal arrows in the top panels, with $\varepsilon_3=5\ueV$.
	Green dot-dashed lines represent average charge in the outer dots, red dashed lines for the inner dots, and the blue solid lines indicate the current.
	Vertical dashed lines mark the position of the resonance.
	The width of the arrows in the top panels indicates the strength of the tunneling rates.
	The parameters for the symmetric case (a, c) are $\tau_i=\left[2, 2, 2\right]\ueV$ and $\varepsilon_1=\varepsilon_4=0$, while for the asymmetric case (b, d) are $\tau_i=\left[2, 1, 3\right]\ueV$, $\varepsilon_1=0$, and $\varepsilon_4=0.4\ueV$.}
	\label{fig:symmetric_vs_asymmetric}
\end{figure}

Fig.~\ref{fig:symmetric_vs_asymmetric} presents the charge current as a function of detunings $\varepsilon_2$ and $\varepsilon_3$ for two distinct configurations.
In the symmetric case(a, c), tunneling amplitudes are equal for all dots, and the outer dots are in resonance.
A pronounced current resonance appears, though the population of the central dots remains non-zero, i.e., $\langle Q_2 \rangle + \langle Q_3 \rangle \neq 0$.
In the case of larger detuning for the inner dots, its population is reduced.
However, the effective tunneling amplitude $\tau_{\text{eff}}$ drastically decreases, leading to a reduced current that is hard to observe in real experiments.

To compare results, we also study an asymmetric configuration (b, d), not only in terms of the on-site energies of the outer dots, but also in the tunneling rates.
In contrast to the symmetric configuration, close to the resonant condition a clear dip in the central dots population emerges.
In this situation, the central dots hybridize effectively forming a single dot.
The resulting dynamics resemble those of a triple dot, which supports a dark state with vanishing central dot population.

Importantly, even in triple quantum dot systems coupled to the leads, where dynamics become incoherent, the central dot population is no longer strictly zero, but still exhibits a pronounced dip close to the resonance, similar to the one shown in Fig.~\ref{fig:symmetric_vs_asymmetric}~(d).
Such a dip is absent in the symmetric QQD configuration.
This observation underscores that long-range transport does not necessarily require a highly symmetric configuration.
Furthermore, the presence of a finite dephasing time $T_2$ modifies both the current and the inner dots' occupation, as can be seen in Appendix~\ref{app:dephasing}.
As expected, systems with low dephasing times exhibit a less pronounced dip in the inner dots' occupation, making it difficult to observe long-range transport.

\begin{figure}[t!]
	\centering
	\includegraphics[width=\linewidth]{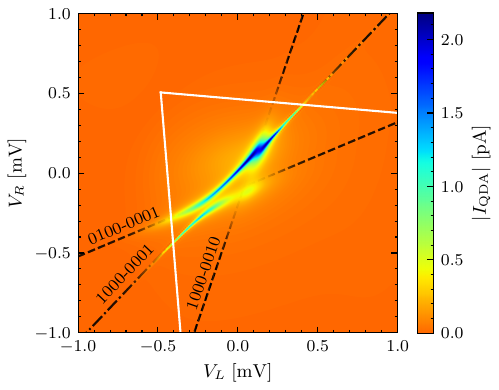}
	\caption{Current through the quadruple quantum dot array in the single-particle regime.
		White lines denote the region of positive charge current, as described in the main text.
		Black lines indicate various resonance conditions.
		Dashed lines correspond to long-range transfers between first and third, and second and fourth dots.
        With a dot-dashed line we denote a long-range transition between the first and last dots.
		Lines fade out in regions of high current to emphasize the current features.
		The parameters are $\tau_i=\left[3,5,2\right]\ueV$, $T_2=10\;\mathrm{ns}$, $\mu_{L(R)} = \pm37.5\ueV$, $\varepsilon_{20}=7\ueV$, and $\varepsilon_{30}=-5\ueV$.}
	\label{fig:one_particle_current}
\end{figure}

So far, we have focused on tuning the on-site energies to gain intuitive insight into the system's behavior.
However, capacitive coupling among the dots significantly alters the resonance conditions.
Nevertheless, based on prior results, we expect that transport signatures persist even under asymmetries induced by lever arms.
In Fig.~\ref{fig:one_particle_current}, we show the charge current as a function of the gate voltages $V_L$ and $V_R$.
Out of the region defined with the white lines, a small but finite current flow due to the finite temperature.
Inside the transport triangle, the current is maximum in a series of resonance conditions, which are represented by the black lines.
These resonances are due to the possibility of a resonant transition between different energy levels of the QQD.
On the one hand, we have the long-range transfer between the first and last dot of the array, labeled as $1000-0001$.
Note that we have drop out the use of kets in the states in order to simplify the notation.
On the other hand, other long-range transitions between the next nearest-neighbor dots, labeled as $0100-0001$, and $1000-0010$, are also present.
When these transitions meet, the current exhibits an avoiding crossing, which is a clear signature of different coherent processes interfering with each other.

Black lines in Fig.~\ref{fig:one_particle_current} represent solutions to the resonance conditions.
As an example, for the $1000-0001$ resonance we solve $\varepsilon_1=\varepsilon_4$, obtaining the condition
\begin{equation}
	V_R = V_L\frac{\alpha_{3L} - \alpha_{0L}}{\alpha_{0R} - \alpha_{3R}}.
	\label{eq:resonance_condition_one_particle}
\end{equation}
By fitting the slope of this resonance, the corresponding lever arms can be extracted.

We conclude that, even in the presence of decoherence and reduced symmetry, manifested through detuned central dots and non-uniform tunneling rates, the $1000-0001$ long-range resonance remains observable.

\section{Multiple particle results}
\subsection{Closed system}
The dynamics of the quadruple quantum dot array become significantly richer when multiple electrons are introduced into the system.
Here, we concentrate on the scenarios involving two and three electrons.

Since the original Hamiltonian described in Eq.~(\ref{eq:hamiltonian_0}) preservers total spin, the Hilbert space decomposes naturally into subspaces labeled by the total spin quantum number $s$.
For two electrons, the total spin can adopt values of $s=1$ (triplet states, $T$) or $s=0$ (singlet states, $S$).
In the case of three electrons, the total spin can adopt values of $s=3/2$ (quartet states, $Q$) or $s=1/2$ (doublet states, $D$).
We employ Clebsch-Gordan coefficients to represent these spin states explicitly:
\begingroup
\renewcommand{\arraystretch}{2}
\begin{table}[h!]
	\centering
	\begin{tabular}{|c|c|c|c|}
		\hline
		Basis & $s$ & $s_z$ & Tensor basis \\
		\hline
		\hline
		$\ket{Q^{(+3 / 2)}_{ijk}}$ & $3 / 2$ & $+3 / 2$ & $\ket{\uparrow_i \uparrow_j \uparrow_k}$ \\
		\hline
		$\ket{Q^{(+1 / 2)}_{ijk}}$ & $3 / 2$ & $+1 / 2$ & $\frac{1}{\sqrt{3}}(\ket{\downarrow_i \uparrow_j \uparrow_k}+\ket{\uparrow_i \downarrow_j \uparrow_k}+\ket{\uparrow_i \uparrow_j \downarrow_k})$ \\
		\hline
		$\ket{Q^{(-1 / 2)}_{ijk}}$ & $3 / 2$ & $-1 / 2$ & $\frac{1}{\sqrt{3}}(\ket{\downarrow_i \downarrow_j \uparrow_k}+\ket{\uparrow_i \downarrow_j \downarrow_k}+\ket{\downarrow_i \uparrow_j \downarrow_k})$ \\
		\hline
		$\ket{Q^{(-3 / 2)}_{ijk}}$ & $3 / 2$ & $-3 / 2$ & $\ket{\downarrow_i \downarrow_j \downarrow_k}$ \\
		\hline
		\hline
		$\ket{D^{(+1 / 2)}_{ijk}}$ & $1 / 2$ & $+1 / 2$ & $\frac{1}{\sqrt{2}}(\ket{\uparrow_i \downarrow_j \uparrow_k}-\ket{\downarrow_i \uparrow_j \uparrow_k})$ \\
		\hline
		$\ket{D^{(-1 / 2)}_{ijk}}$ & $1 / 2$ & $-1 / 2$ & $\frac{1}{\sqrt{2}}(\ket{\uparrow_i \downarrow_j \downarrow_k}-\ket{\downarrow_i \uparrow_j \downarrow_k})$ \\
		\hline
		\hline
		$\ket{D'^{(+1 / 2)}_{ijk}}$ & $1 / 2$ & $+1 / 2$ & $\frac{1}{\sqrt{6}}(2\ket{\uparrow _i\uparrow_j \downarrow_k}-\ket{\uparrow_i \downarrow_j \uparrow_k}-\ket{\downarrow_i \uparrow_j \uparrow_k})$ \\
		\hline
		$\ket{D'^{(-1 / 2)}_{ijk}}$ & $1 / 2$ & $-1 / 2$ & $-\frac{1}{\sqrt{6}}(2\ket{\downarrow_i \downarrow_j \uparrow_k}-\ket{\downarrow_i \uparrow_j \downarrow_k}-\ket{\uparrow_i \downarrow_j \downarrow_k})$ \\
		\hline
	\end{tabular}
\end{table}
\endgroup
Here, indices $i \neq j \neq k \neq i$ indicate distinct quantum dots.
In addition to these states, doubly occupied states, denoted by $\ket{S_i \sigma_j}$, indicating a double occupancy at site $i$ and single occupancy with spin $\sigma$ at site $i\neq j$, also become relevant.

The system's dynamics are thus enriched by competing coherent processes.
To elucidate the underlying physics, it is instructive to analyze each spin sector independently.
The simplest scenario is the $s=3/2$ subspace, where double occupancies are forbidden.
It is worth marking that no spin-flip processes are present in the closed system, so no coupling between quartets and double-occupied state are possible.
Under these conditions, the sector closely resembles the single-particle, admitting long-range transfer when $E_{1110}=E_{0111}$.
In such conditions, the central dots are maximally occupied, yet exhibit no charge fluctuations during dynamics.
Interestingly, this resonance condition coincides exactly with the single-particle resonance given by Eq.~(\ref{eq:resonance_condition_one_particle}).
The condition for positive current in this regime is $\mu_L > (E_{1110} - E_{0110})$ and $\mu_R < (E_{0111} - E_{0110})$.
Near resonance, we derive an effective two-state Hamiltonian effective model given by:
\begin{align}
	\begin{split}
		\hat{H}_\mathrm{eff} = & \tilde{E}_{1110}\ketbra{Q^{(+3/2)}_{123}}{Q^{(+3/2)}_{123}}+\tilde{E}_{0111}\ketbra{Q^{(+3/2)}_{234}}{Q^{(+3/2)}_{234}} \\
		                       & + \tilde{\tau}\left(\ketbra{Q^{(+3/2)}_{123}}{Q^{(+3/2)}_{234}}+\mathrm{h.c.}\right),
	\end{split}
	\label{eq:effective_model_1110}
\end{align}
where the effective energies and hopping amplitudes can be found in Appendix~\ref{app:effective_model_1110-0111}.
By making the substitutions $E_{1110}\rightarrow\varepsilon_4$, $E_{1101}\rightarrow\varepsilon_3$, $E_{1011}\rightarrow\varepsilon_2$, and $E_{0111}\rightarrow\varepsilon_1$, we directly mapped to a single-particle scenario by reinterpreting the process as hole transport between the outermost dots.
In the equation above we have focused on the $s_z=+3/2$ sector, but the same results can be obtained for all the other $s_z$ values.

In the $s=1/2$ subspace, analogous long-range transfer conditions ($E_{1110}=E_{0111}$) remain valid.
However, due to degeneracy between doublet states $\ket{D_{ijk}^{(\sigma)}}$ and $\ket{D_{ijk}'^{(\sigma)}}$ in the absence of magnetic fields, the resulting effective model is significantly more intricate, making the mapping to a single-particle systems less straightforward.

Alternatively, double occupancy states permit another type of long-range transport under the resonance $E_{2001}=E_{1002}$, characterized by minimal intermediate-dot occupancy.
An effective Hamiltonian describing this case reads
\begin{align}
	\begin{split}
		\hat{H}_\mathrm{eff} = & \tilde{E}_{2001}\ketbra{S_1\downarrow_4}{S_1\downarrow_4}+\tilde{E}_{1002}\ketbra{S_4\downarrow_1}{S_4\downarrow_1} \\
		                       & + \tilde{\tau}\left(\ketbra{S_1\downarrow_4}{S_4\downarrow_1}+\mathrm{h.c.}\right),
	\end{split}
	\label{eq:effective_model_2001}
\end{align}
where the effective energies and hopping amplitudes can be found in Appendix~\ref{app:effective_model_2001-1002}.
Fig.~\ref{fig:effective_model_three_particles} shows the dynamics of the closed system, which demonstrates that this effective model accurately captures the system's dynamics.

\begin{figure}[t!]
	\centering
	\includegraphics[width=\linewidth]{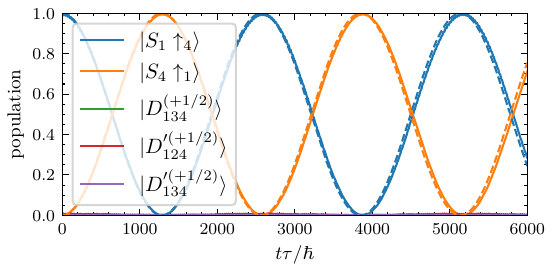}
	\caption{Dynamics (solid lines) for three particles in a quadruple quantum dot array.
		The effective model from Eq.~(\ref{eq:effective_model_2001}) shown with dashed lines.
		System parameters are given by $\tau_i = \tau=3 \ueV$, $\varepsilon_i=\left[-1836.2, -700, -655, -1714\right] \ueV$, $U_i = \left[1720, 1500, 1220, 1600\right] \ueV$, $V_i=\left[450, 200, 100\right] \ueV$.}
	\label{fig:effective_model_three_particles}
\end{figure}

Here, we have shown the dynamics for the case of three particles, but similar results can be obtained for two particles for the long-range resonances $2000-1001$ and $1100-0101$.

\subsection{Open system} \label{sec:open_system_three_particles}
Coupling the quadruple quantum dot to leads enables a finite current to flow under suitable resonance conditions.
A general resonance between two charge configurations, $\ket{n_1,n_2,n_3,n_4}$ and $\ket{n_1',n_2',n_3', n_4'}$, is described by a straight line in the $V_L$-$V_R$ plane.
Solving the resonance condition $E_{n_1,n_2,n_3,n_4}=E_{n_1',n_2',n_3', n_4'}$, the resulting line takes the form $V_R = m V_L + n$, where the slop is given by
\begin{equation}
	m = \dfrac{\sum_{i=1}^4 \alpha_{iL}(n_i - n_i')}{\sum_{i=1}^4 \alpha_{iR}(n_i' - n_i)}.
\end{equation}
Note that, in general, the slope $m$ is not unique, and different resonances may exhibit the same slope.
For example, the transitions $2000-1001$ and $1100-0101$ share the same value of $m$.
The intercept $n$ is determined by the Coulomb interaction and the on-site energies $\varepsilon_{20}$ and $\varepsilon_{30}$, defined in Eq.~(\ref{eq:lever_arms}).
It is this intercept that breaks the slope degeneracy, allowing us to distinguish between different resonance lines.
However, since the explicit expression for $n$ is lengthy and not relevant for the current discussion, we omit it here.

\begin{figure}[t!]
	\centering
	\includegraphics[width=\linewidth]{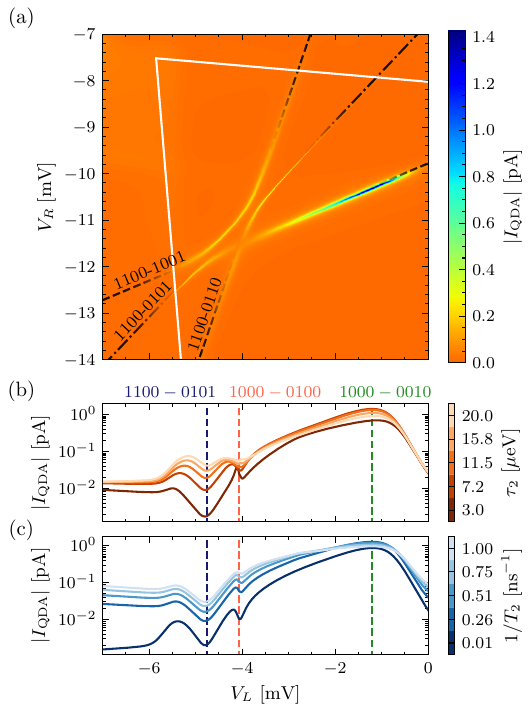}
	\caption{(a) Current through the quadruple quantum dot array in the two-particle regime.
		White lines indicate the region of positive current, as defined in the main text.
		Black dashed and dot-dashed lines correspond to different resonance conditions, as labeled.
		(b-c) Current blockade for the $1100-1001$ resonance.
		In (b) we show the current for different values of the central tunneling amplitude $\tau_2$ for $T_2 = 10$ ns, and in (c) we show the current for different values of the dephasing time $T_2$ with $\tau_2 = 5\ueV$.
		Vertical color-coded dashed lines indicate the position for the resonances denoted on top of the panel (b).
		Other parameters, common to both panels, are $\varepsilon_{20}=8\ueV$ and $\varepsilon_{30}=10\ueV$, $U_i = \left[1500, 1650, 1220, 1800\right] \ueV$, $V_i=\left[1000, 500, 333\right] \ueV$, $\tau_i=\left[10, 20, 10\right]\ueV$, and $\mu_L = 1550\ueV$ and $\mu_R=1250\ueV$.
		All other parameters are identical to those in Fig.~\ref{fig:one_particle_current}.}
	\label{fig:two_particles_1100-0101_blockade}
\end{figure}

In Fig.~\ref{fig:two_particles_1100-0101_blockade}~(a), we show the charge current near the $1100-0101$ transition.
In this case, both singlet and triplet states contribute to the transport.
Since we are dealing with multiple resonance conditions with different initial and final states, the transport triangle differs for each resonance condition.
This difference is clearly visible in the $1100-1001$ resonance, where the current rapidly decreases for $V_L>-2$ mV.
Across the entire work, we will denote the region of positive current with white lines for the long-range transfer between the first and last dots.
Other regions of positive current can be obtained by imposing the condition of positive current.
Interestingly, at the intersection of the $1100-0110$ and $1100-0101$ resonances, we observe a pronounced current drop.

This current blockade is a consequence of the resonance between the first and third dots.
To gain better insight into the current blockade, we focus on the $1100-1001$ resonance.
For this resonance, out of the blockade region, the current follows the path $1000\rightarrow 0100 \rightarrow 1100 \rightarrow 1001 \rightarrow 1000$.
The first step is the bottleneck of the process, since other paths are in resonance.
That is why, the first and the second dots are in resonance, the current is maximized, see Fig.~\ref{fig:two_particles_1100-0101_blockade}~(b-c).
However, when the first and the third dots are in resonance, the path is given by $1000\rightarrow 0010 \rightarrow 1010$.
The final state is out of the transport window, and the current is blockaded, leading a sharp drop in the current.
Since double occupancy states are not involved in the blockade, the same mechanism explains the blockade of triplet and singlet states.
If the tunneling rate between the second and the third dots is increased, even if there are out of resonance, a finite current can flow through the system, obtaining a less pronounced blockade, as shown in Fig.~\ref{fig:two_particles_1100-0101_blockade}~(b).
When the dephasing time is increased, see Fig.~\ref{fig:two_particles_1100-0101_blockade}~(c), the current blockade is slightly lifted, but a clear current drop is still present.
It is important to note that the blockade also coincide with the resonance $E_{1100}=E_{1001}=E_{0110}$.
However, the presence of the state $\ket{0110}$ is not necessary for the blockade to occur.
This coincidence is due to the fact that the interdot Coulomb repulsion is site independent, i.e, the repulsion energy between two particles located between the first and the second dots is the same as between the second and the third dots.
The result is that the resonant condition for the $1100-0110$ transition is given by the same condition as the $1000-0010$ transition.

\begin{figure}[t!]
	\centering
	\includegraphics[width=\linewidth]{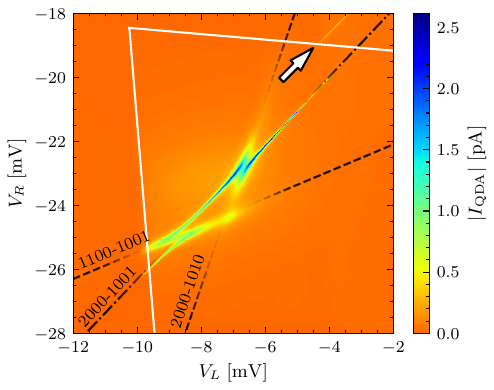}
	\caption{Current through the quadruple quantum dot array in the two-particle regime.
		White lines indicate the region of positive current, as defined in the main text.
		Black dashed and dot-dashed lines correspond to different resonance conditions, as labeled.
		The detuning of the central dots is $\varepsilon_{20}=800\ueV$ and $\varepsilon_{30}=9203.7\ueV$, and $\mu_L = 2500\ueV$ and $\mu_R=1900\ueV$.
		A white arrow indicates the $2010-1101$ resonance.
		All other parameters are identical to those in Fig.~\ref{fig:two_particles_1100-0101_blockade}.}
	\label{fig:two_particles_2000-1001}
\end{figure}

By modifying the bias voltage, and the detuning of the central dots, we can focus on resonance conditions such that singlet states are mainly involved in the transport while reducing the presence of triplet states, as shown in Fig.~\ref{fig:two_particles_2000-1001}.
Here, only the resonance condition $1100-1001$ carry any contribution from the triplet states, as demonstrated by the absence of an avoided crossing between the $2000-1001$ and $1100-0101$ transitions.
In the upper right corner, marked with a white arrow, we observe a fine line with high current which corresponds to a three-particle resonance, given by $2010-1101$.
This resonance is interesting since it corresponds to a tunneling involving two electrons, in contrast to the previous cases, where only one electron was involved in the tunneling process.

\begin{figure}[t!]
	\centering
	\includegraphics[width=\linewidth]{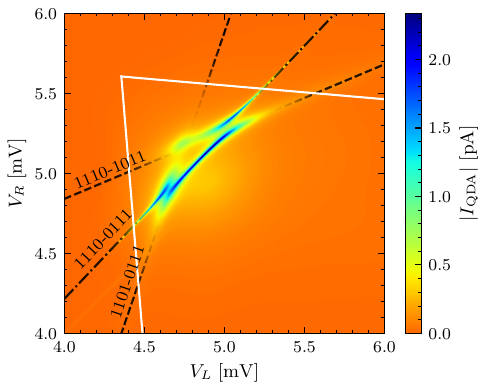}
	\caption{Current through the quadruple quantum dot array in the three-particle regime.
		White lines indicate the region of positive current, as defined in the main text.
		Black dashed and dot-dashed lines correspond to different resonance conditions, as labeled.
		The detuning of the central dots is $\varepsilon_{20}=\varepsilon_{30}=-398\ueV$, and the Coulomb repulsion is $U_i = \left[1720, 1500, 1220, 1600\right] \ueV$, $V_i=\left[280, 167, 100\right] \ueV$.
		All other parameters are identical to those in Fig.~\ref{fig:one_particle_current}.}
	\label{fig:three_particle_current_1110-0111}
\end{figure}

Going beyond the two-particle regime, more interesting phenomena can be found when including an extra particle as show in Fig.~\ref{fig:three_particle_current_1110-0111}, where we show the charge current near the $1110-0111$ transition.
Here, both quartet and doublet states contribute to the transport.
Although there are no doubly occupied states involved, long-range transfer remains possible.
The results bear a resemblance to the single-particle case, as seen by comparing with Fig.~\ref{fig:one_particle_current} after applying the transformation $V_{L(R)} \rightarrow - V_{R(L)}$.
This symmetry arises from the interpretation of the $1110-0111$ resonance as a long-range hole transfer, equivalent to a single positively charged particle moving from right to left.

The resonance condition for the $2001-1002$ transition is given by
\begin{equation}
	V_R = \frac{V_L(\alpha_{4L} - \alpha_{1L})+U_4-U_1}{\alpha_{1R} - \alpha_{4R}},
	\label{eq:resonance_condition_2001-1002}
\end{equation}
while the condition for positive current is $\mu_L > (E_{2001} - E_{1001})$ and $\mu_R < (E_{1002} - E_{1001})$.
However, one must ensure that this transport window does not overlap with the regions where a fourth electron can enter the system, which would significantly suppress the current.
To avoid this, we impose the additional constraints $\mu_L < (E_{1111} - E_{0111})$ and $\mu_R > (E_{1110} - E_{1111})$.
Although the resonance condition is independent of $\mu_{L(R)}$, the positive current region depends explicitly on the chemical potentials of the leads.
These can be parametrized as $\mu_{L(R)} = \mu_0 \pm \Delta \mu /2$.
With this formulation, $\mu_0$ shifts the transport triangle across the $V_L=V_R$ line, while $\Delta \mu$ displaces it along the $V_L=-V_R$ direction.

To satisfy the above constraints while avoiding the addition of an extra particle, the system parameters must be selected carefully, with particular attention to the interdot Coulomb interactions $V_i$.
We find that, for fixed values of all other parameters, there exists only a narrow window of $V_i$ where a positive current and visible long-range transfer coexist.
This regime typically occurs for large values of $V_i$.

\begin{figure}[t!]
	\centering
	\includegraphics[width=\linewidth]{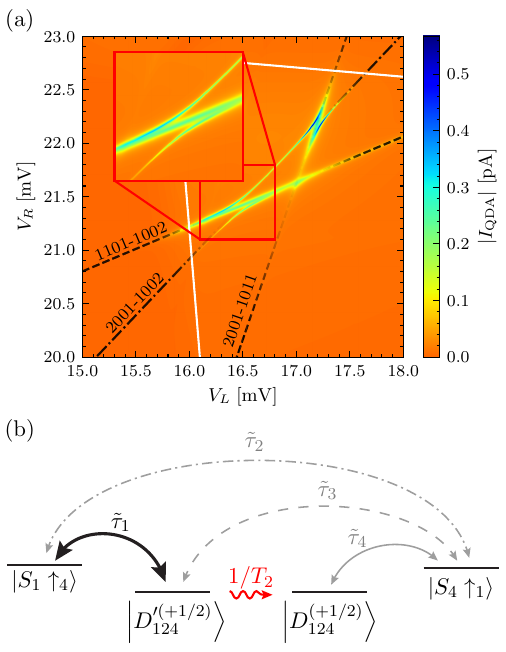}
	\caption{(a) Current through the quadruple quantum dot array in three-particles region.
		White lines indicate the region of positive current, defined in the main text.
		Black dashed and dot-dashed lines represent different resonance conditions, as labeled.
		The detuning of the center dots is $\varepsilon_{20}=\varepsilon_{30}=-367\ueV$, and $U_i = \left[1500, 1650, 1220, 1800\right] \ueV$, $V_i=\left[1000, 500, 333\right] \ueV$.
        (b) Schematic of the effective model for the $2001-1002-1101$ resonance.
        Strong coupling is depicted with a thick black line, while weak couplings are represented with thin gray lines.
        A dephasing mechanism is represented by a red wavy line.}
	\label{fig:three_particle_current_2001-1002}
\end{figure}

Fig.~\ref{fig:three_particle_current_2001-1002}~(a) displays the charge current as a function of the gate voltages $V_L$ and $V_R$ for a representative set of parameters.
The chosen configuration is one of many that yields a high-current region associated with long-range transfer.
The resulting current is highly similar to the two-particle case shown in Fig.~\ref{fig:two_particles_2000-1001}.
This similarity arises since the particle stuck in the last dot does not affect the dynamics of the system, and only renormalizes the different energies.
Then, this particle can be traced out, leading to an effective model with only two particles.
Due to the complexity of the system, multiple parameter combinations can produce similar results.
Here, we observe three distinct long-range transitions, $2001-1002$, $1110-1002$, and $2001-1011$, each indicated by black lines.
At the intersections of these transitions, the current exhibits avoided crossings, which signal interference between coherent transport processes.
This is particularly clear in the upper right region, where the $2001-1002$ and $2001-1011$ cross.

In contrast, at the crossing between $2001-1002$ and $1101-1002$, shown in the zoomed view in Fig.~\ref{fig:three_particle_current_2001-1002}~(a), a straight unperturbed resonance line appears.
To understand this behavior, we analyze the effective model near the triple degeneracy point $E_{2001}=E_{1002}=E_{1101}$.
The corresponding four-level Hamiltonian, expressed in the basis $\left\{\ket{S_1\uparrow_4},\ket{D_{124}'^{(+1/2)}},\ket{D_{124}^{(+1/2)}}, \ket{S_4\uparrow_1} \right\}$, reads
\begin{equation}
	\hat{H}_\mathrm{eff} = \mqty(E_1 & \tilde{\tau}_1 & 0 & \tilde{\tau}_2 \\
	\tilde{\tau}_1 & E_2 & 0 & \tilde{\tau}_3\\
	0 & 0 & E_3 & \tilde{\tau}_4\\
	\tilde{\tau}_2 & \tilde{\tau}_3 & \tilde{\tau}_4 & E_4).
	\label{eq:effective_model_2001-1002}
\end{equation}
Explicit expressions for the effective energies and hopping amplitudes are provided in Appendix~\ref{app:effective_model_2001-1002}.
Using the chosen working parameters, we find that $\abs{\tilde{\tau}_1}\gg \abs{\tilde{\tau_2}}, \abs{\tilde{\tau}_3}, \abs{\tilde{\tau}_4}$.
Diagonalizing the Hamiltonian reveals two pairs of eigenstates, in which $\ket{S_1\uparrow_4}$ and $\ket{D_{124}'^{(+1/2)}}$ weakly coupled to the remaining states.
Since $\ket{S_4\uparrow_1}$ is the only state within the transport window, current is suppressed, and the system becomes trapped in the first pair of states, leading to current blockade.
Nonetheless, a dephasing mechanism can induce leakage from the first pair of states to the second one enabling a finite current.
In the central region of the crossing in Fig.~\ref{fig:three_particle_current_2001-1002}~(a), the current flows via the sequence $\ket{S_1\uparrow_4}\rightarrow\ket{D_{124}'^{(+1/2)}}\rightarrow\ket{D_{124}^{(+1/2)}}\rightarrow\ket{S_4\uparrow_1}$.
Although $\ket{D_{124}'^{(+1/2)}}$ and $\ket{D_{124}^{(+1/2)}}$ are not directly coupled in the effective Hamiltonian, dephasing enables transitions between them.
This process is schematically illustrated in Fig.~\ref{fig:three_particle_current_2001-1002}~(b).
In the limit of vanishing dephasing ($T_2\rightarrow \infty$), not shown here, the current in the central region of the anticrossing disappear.

\section{Conclusions}
We have analyzed long-range transfer in a quadruple quantum dot array under conditions of reduced symmetry.
First, we investigated the single-particle regime.
Using experimentally feasible parameters, we found that direct transfer between the first and last dots remains observable even in the presence of decoherence.
An avoided crossing in the charge current, arising between distinct long-range processes, serves as a clear signature of coherent interference.
We derived an effective model that accurately captures the system' dynamics, and obtained a simple analytical expression for the resonance condition associated with long-range transfer.

We further explored how long-range transfer persists in an interacting system with two and three electrons in the QQD.
In this regime, we analyzed the contributions of different spin sectors independently, finding visible long-range transfers in all cases.
In the case of two electrons, we find a coherent spin blockade of the current, which survives even in the presence of dephasing.
For three particles in the $s=3/2$ sector, we demonstrated that long-range transfer remains possible even when the central dots are singly occupied, yielding results analogous to the single-particle case.
For the $s=1/2$ sector, we showed that the two distant doublet states, though degenerate in absence of magnetic fields, can be distinguished through the influence of dephasing, which imprints a characteristic signature in the charge current.

\section{Acknowledgements}
G.P. and D.F.F. are supported by the Spanish Ministry of Science through the grant: PID2023-149072NB-I00 and by the CSIC Research Platform PTI-001.
D.F.F. acknowledges support from FPU Program No. FPU20/04762.
J.C.B and R.J.H are supported by the Deutsche Forschungsgemeinschaft (DFG, German Research Foundation) under Germany's Excellence Strategy - EXC 2123 QuantumFrontiers - 390837967 and the State of Lower Saxony of Germany via the Hannover School for Nanotechnology.

\appendix
\section{Derivation of the Lindblad master equation}\label{app:master_eq}
The master equation, within the Born-Markov approximation, is given by \cite{Busl2013}
\begin{equation}
	\begin{split}
		\dot{\rho}_{m n}(t)= & -i\left\langle m\left|\left[\hat{H}_0, \hat{\rho}\right]\right| n\right\rangle            \\
		                     & +\sum_{k \neq n}\left(\Gamma_{n k} \rho_{k k}-\Gamma_{k n} \rho_{n n}\right) \delta_{m n} \\
		                     & -\Lambda_{m n} \rho_{m n}\left(1-\delta_{m n}\right),
	\end{split}
\end{equation}
where the first term on the right-hand side is the coherent evolution of the density matrix, while the second and third terms account for the incoherent processes.
Transition rates $\Gamma_{m,n}$ between states $m$ and $n$ are of two kinds.
The first one is sequential tunneling through the leads, which is given by Fermi's golden rule
\begin{align}
	\begin{split}
		\Gamma_{m n}=\sum_{l=\mathrm{L}, \mathrm{R}} \Gamma_l & \left\{f\left(E_m-E_n-\mu_l\right) \delta_{N_m, N_n+1}\right.                  \\
		                                                      & \left.+\left[1-f\left(E_n-E_m-\mu_l\right)\right] \delta_{N_m, N_n-1}\right\},
	\end{split}
\end{align}
where $E_m$ is the energy of the state $\ket{m}$ of an isolated quantum dot array, $\mu_{L, R}$ is the chemical potential of the left and right leads, respectively, $f(E)$ is the Fermi-Dirac distribution function, and $\Gamma_{L, R}=2\pi\mathcal{D}_{L, R}\abs{\gamma_{L, R}}^2$ is the tunneling rate for each lead, with $\mathcal{D}_{L, R}$ the density of states of the leads.

The second kind of incoherent transition rate is due to the relaxation processes, which are modeled as a phenomenological spin-flip rate due to hyperfine interaction.
The spin relaxation time $T_1$ is given by $T_1^{-1} = (W_{\uparrow\downarrow}+W_{\downarrow\uparrow})$, where $W_{\uparrow\downarrow}$ and $W_{\downarrow\uparrow}$ fulfill a detailed balance condition $W_{\downarrow\uparrow}^i/W_{\uparrow\downarrow}^i = e^{-\beta\Delta_i}$, with $\Delta_i$ the effective Zeeman splitting in the $i$-th quantum dot, and $\beta=1/k_BT$.

Finally, the dephasing rate $\Lambda_{m n}$ is given by
\begin{equation}
	\Lambda_{mn} = \frac{1}{2}\left(\sum_{k\neq m}\Gamma_{km} + \sum_{k\neq n}\Gamma_{kn}\right) + \frac{1}{T_2},
\end{equation}
where $T_2$ is the dephasing time.

During all the calculations, we have assumed an isotropic hyperfine interaction, with values $\Delta_i = \left[0.15, 0.2, 0.1, 0.14\right] \ueV$.
The main conclusions of the paper are independent of the specific values of the hyperfine interaction, as long as they are in the range of typical values for GaAs quantum dots.
Finally, we also have considered a temperature of $T=0.1$ K, a spin relaxation time $T_1=1\;\mathrm{ms}$, and the coupling to the leads $\Gamma_L = 0.28\ueV$, $\Gamma_R = 0.21\ueV$.

\section{Derivation of the current operator}\label{app:current_op}
To compute the current operator, we assume a positive bias direction, i.e., $\mu_L > \mu_R$.
Therefore, the current operator is given by
\begin{equation}
	\hat{I} = e\sum_{m,n}\left(\Gamma_{mn}^+\rho_{nn} - \Gamma_{mn}^-\rho_{nn}\right),
\end{equation}
where $\Gamma_{mn}^+$ is the transition rate from state $\ket{m}$ to state $\ket{n}$ due to a tunneling event from the quantum dot array to one lead, and analogous for $\Gamma_{mn}^-$ for a tunneling event from one lead to the quantum dot array.

\section{Effect of dephasing}\label{app:dephasing}
The effect of dephasing in the system is hard to analyze, since it depends on the specific parameters of the system.
For instance, in Section~\ref{sec:open_system_three_particles}, we have shown that the presence of a dephasing mechanism can reveal a long-range resonance between the $2001-1002$ and $1101-1002$ transition, which is not present in the case of no dephasing.
In this appendix, we show the effect of dephasing in the case of a single particle in the system, in the asymmetric configuration, as the one shown in Fig.~\ref{fig:symmetric_vs_asymmetric}~(c-d).

\begin{figure}[t!]
	\centering
	\includegraphics[width=\linewidth]{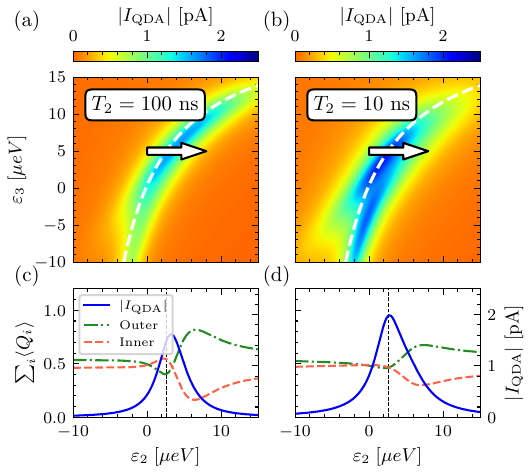}
	\caption{Comparison of current for a high dephasing time $T_2=100\;\mathrm{ns}$ (a) and a low dephasing time $T_2=10\;\mathrm{ns}$ (b).
	White dashed lines denote the analytical resonance condition from Eq.~(\ref{eq:resonance_condition_one_particle_detuning}).
	Panels (c, d) show the average charge (left axis) and current (right axis) along the white horizontal arrows in the top panels.
	Green dot-dashed lines represent average charge in the outer dots, red dashed lines for the inner dots, and the blue solid lines indicate the current.
	Vertical dashed lines mark the position of the resonance.
	The parameters for both cases are $\tau_i=\left[2, 1, 3\right]\ueV$, $\varepsilon_1=0$, and $\varepsilon_4=0.4\ueV$.}
	\label{fig:low_dephasing_time}
\end{figure}

In Fig.~\ref{fig:low_dephasing_time} we show the current for a high dephasing time $T_2=100\;\mathrm{ns}$ (a), and a low dephasing time $T_2=10\;\mathrm{ns}$ (b).
The general behavior of the current is similar in both cases, with a small current increase in the resonance region for the low dephasing time.
Furthermore, while preserving the existence of a dip, it can be seen that decreasing the dephasing time increases the population of the inner dots.
In the case of very low dephasing time, e.g., $T_2\sim 1$ ns, the resonance condition is distorted, and the dip in the population of the inner dots is not present.
We have also checked the opposite limit, in which no dephasing is present.
In this case, a high current is still present, and the dip obtain a value close to $\langle Q_2\rangle + \langle Q_3 \rangle \sim 0.1$.

\begin{figure}[t!]
	\centering
	\includegraphics[width=\linewidth]{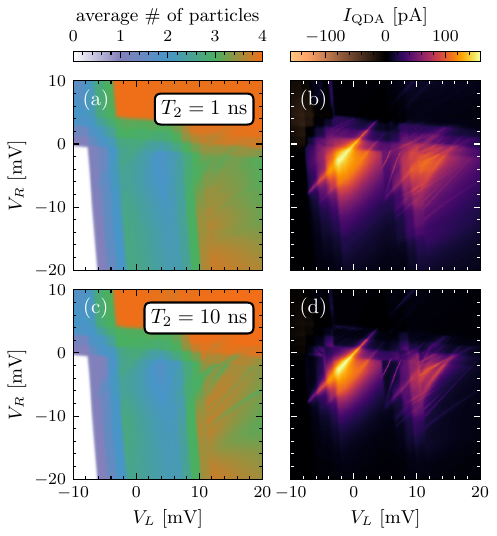}
	\caption{(a, c) Average number of particles in the system, and (b, d) current through the quadruple quantum dot array versus the left and right gate voltages.
	The upper panels (a, b) correspond to a low dephasing time $T_2=0.1\;\mathrm{ns}$, while the lower panels (c, d) correspond to a high dephasing time $T_2=10\;\mathrm{ns}$.
	Other parameters, common to all panels, are: $\tau_i = 100\ueV$, $\varepsilon_{i0} = 0$, $U_i=1$ meV, $V_i = \left[250, 125, 50\right] \ueV$, $\Gamma_i = 1\ueV$, and $\mu_L=656\ueV$, $\mu_R=81\ueV$.}
	\label{fig:dephasing_effect_multi_particle}
\end{figure}

When working in the high tunneling regime, the dephasing time has an even stronger impact.
In Fig.~\ref{fig:dephasing_effect_multi_particle} we show the effect of dephasing in the system, for a high dephasing time $T_2=1\;\mathrm{ns}$ (a, b), and a low dephasing time $T_2=10\;\mathrm{ns}$ (c-d).
Here, we can clearly see that, in presence of low dephasing time, the spin blockade is lifted, resulting in a current leakage and hiding other resonances in the system.
Furthermore, even a negative current is present, as can be seen in the upper left corner of Fig.~\ref{fig:dephasing_effect_multi_particle}~(b).
When increasing the dephasing time, the spin blockade is restored, and the long-range transitions are visible again.
These results show the importance of having a high enough dephasing time in the system, in order to observe the long-range transfer.

\section{Effective model for the \texorpdfstring{$1110-0111$}{1110-0111} transition}\label{app:effective_model_1110-0111}
The effective model for the $1110-0111$ transition, is given by the Hamiltonian in Eq.~(\ref{eq:effective_model_1110}).
The effective energies and hopping amplitudes can be obtained by a Schrieffer-Wolff transformation, which results in
\begin{subequations}
	\begin{equation}
		\tilde{E}_{1110}\equiv E_{1110} + \frac{\tau_3^2}{E_{1110}-E_{1101}},
	\end{equation}
	\begin{equation}
		\tilde{E}_{0111}\equiv E_{0111} + \frac{\tau_1^2}{E_{0111}-E_{1011}}
	\end{equation}
	\begin{equation}
		\tilde{\tau}\equiv\frac{\tau_1\tau_2\tau_3}{(E_{1110}-E_{1101})(E_{0111}-E_{1011})}.
	\end{equation}
\end{subequations}

\section{Effective model for the \texorpdfstring{$2001-1002$}{2001-1002} transition}\label{app:effective_model_2001-1002}
The effective model for the $2001-1002$ transition, far from other resonances, is given in Eq.~\ref{eq:effective_model_2001}.
The effective energies and hopping amplitudes can be obtained by a Schrieffer-Wolff transformation, which results in
\begin{subequations}
	\begin{equation}
		\tilde{E}_{2001}\equiv E_{2001} + \frac{2 \tau_1^2}{E_{2001}-E_{1101}}+\frac{\tau_3^2}{E_{2001}-E_{2010}},
	\end{equation}
	\begin{equation}
		\tilde{E}_{1002}\equiv E_{1002} + \frac{2 \tau_3^2}{E_{1002}-E_{1011}}+\frac{\tau_1^2}{E_{1002}-E_{0102}},
	\end{equation}
	\begin{align}
		\begin{split}
			\tilde{\tau}\equiv\frac{\tau_1\tau_2\tau_3}{2} & \left[(E_{1002}-E_{1011})^{-1}
			(E_{1002}-E_{1101})^{-1}\right.                                                                                    \\
			                                                & \left.+(E_{2001}-E_{1011})^{-1} (E_{2001}-E_{1101})^{-1}\right].
		\end{split}
	\end{align}
\end{subequations}

Close to the resonance condition $E_{2001}=E_{1002}=E_{1101}$, is given by the Hamiltonian in Eq.~(\ref{eq:effective_model_2001-1002}), with the effective energies and hopping amplitudes given by
\begin{subequations}
	\begin{equation}
		E_1 = E_{2001} + \frac{\tau_3^2}{E_{2001} - E_{2010}},
	\end{equation}
	\begin{equation}
		\begin{split}
			E_2 = & E_{1101} + \frac{2\tau_1^2}{E_{2001}-E_{0201}}+\frac{\tau_2^2}{E_{2001} - E_{1011}} \\
			      & + \frac{\tau_3^2}{E_{2001} - E_{1110}},
		\end{split}
	\end{equation}
	\begin{equation}
		E_3 = E_{1101} + \frac{\tau_2^2}{E_{2001} - E_{1011}} + \frac{\tau_3^2}{E_{2001} - E_{1110}},
	\end{equation}
	\begin{equation}
		E_4 = E_{1002} + \frac{\tau_1^2}{E_{2001} - E_{0102}} + \frac{2\tau_3^2}{E_{2001}-E_{1011}},
	\end{equation}
\end{subequations}

\begin{subequations}
	\begin{align}
		\begin{split}
			\tilde{\tau}_1 = \frac{\tau_1}{\sqrt{2}} & \left[\frac{2 \tau_1^2}{(E_{0201}-E_{2001})^2}+\frac{\tau_2^2}{(E_{1011}-E_{2001})^2}\right.        \\
			                                         & \left.+\frac{\tau_3^2 (E_{1110}-E_{2010})^2}{(E_{1110}-E_{2001})^2 (E_{2001}-E_{2010})^2}-2\right],
		\end{split}
	\end{align}
	\begin{equation}
		\tilde{\tau}_2 = \frac{\tau_1 \tau_2 \tau_3}{2(E_{1011}-E_{2001})^2},
	\end{equation}
	\begin{equation}
		\tilde{\tau}_3 = \frac{\tau_2\tau_3}{\sqrt{2} (E_{2001}-E_{1011})},
	\end{equation}
	\begin{equation}
		\tilde{\tau}_4 = \sqrt{\frac{3}{2}}\frac{\tau_2\tau_3}{E_{1011}-E_{2001}}.
	\end{equation}
\end{subequations}

\section{Eigenenergies}\label{app:eigenenergies}
\begin{figure}[h!]
	\centering
	\includegraphics[width=\linewidth]{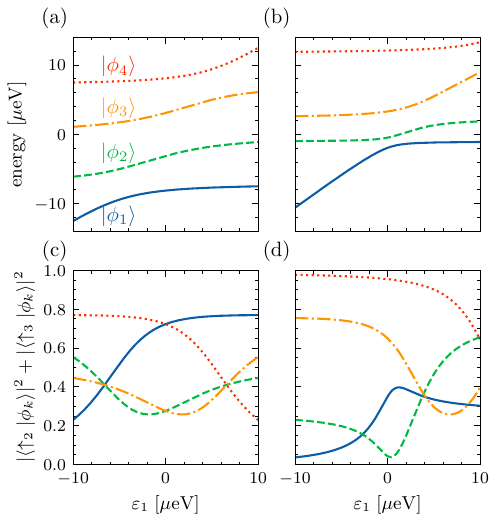}
	\caption{Eigenenergies of the closed system with a single particle as a function of the detuning $\varepsilon_1$ for a symmetric configuration (a), and an asymmetric configuration (b).
	Lower panels show the contribution from the middle dots in each eigenstate $\ket{\phi_k}$, for the symmetric (c) and asymmetric (d) configurations.
	The parameters for the symmetric configuration are $\tau_i = 5\ueV$, $\varepsilon_i = 0$, and for the asymmetric configuration $\tau_i = \left[3, 5, 2\right]\ueV$, $\varepsilon_2=7\ueV$, $\varepsilon_3=6\ueV$ and $\varepsilon_4=0$.}
	\label{fig:band_structure_one_particle}
\end{figure}

In order to better understand the contribution of the different states to the current in the quadruple quantum dot, we analyze the eigenenergies of the system, for the different number of particles.
In Fig.~\ref{fig:band_structure_one_particle} we show the eigenenergies of the system for a symmetric configuration (a), and an asymmetric configuration (b).
In this case, the eigenstate that is responsible for a non-zero current is $\ket{\phi_2}$ (green dashed line).
Looking at the population of the middle dots in this eigenstates in the symmetric configuration Fig.~\ref{fig:band_structure_one_particle}~(c), when all dots are in resonance with $\varepsilon_i=0$, the transfer is sequential, and the population of the middle dots is high.
However, when the middle dots are detuned, Fig.~\ref{fig:band_structure_one_particle}~(d), the population of the middle dots close to the resonance between the outermost dots drastically decreases, with a clear dip.
The position of the dip is well described by the effective model for a single particle given in Eq.~(\ref{eq:effective_model}).
The slight shift from the condition $\varepsilon_1 = \varepsilon_4 = 0$ is due to the mismatch of the tunneling rates $\tau_1\neq \tau_3$ and $\varepsilon_3 \neq \varepsilon_4$.

In the limit of high detuning between the center dots and the outer dots, and effective middle dot is obtained, giving raise to an effective triple quantum dot, given by the three lower energies in Fig.~\ref{fig:band_structure_one_particle}.
In this case, a dark state is formed, given by $\ket{DS} = \mathcal{N}(\ket{\uparrow_1}+\eta\ket{\uparrow_4})$, where $\mathcal{N}$ is a normalization factor, and $\eta$ is a function of the ratio between the tunneling rates $\tau_1$ and $\tau_3$.
The contribution of the middle dots can be added perturbatively.
However, it is negligible, as can be seen in Fig.~\ref{fig:band_structure_one_particle}~(d).

\newpage

\bibliography{references.bib}

\end{document}